# Evolution of superconductivity in ultrathin NbS$_2$

*Rusen Yan,[1,a] Guru Khalsa,[2,3] Brian T. Schaefer,[4] Alexander Jarjour,[4] Sergei Rouvimov,[5]*

*Katja C. Nowack,[4,6] Huili G. Xing,[1,2,6] and Debdeep Jena[1,2,b]*

1. School of Electrical and Computer Engineering, Cornell University, Ithaca, NY 14853, USA

2. Department of Materials Science and Engineering, Cornell University, Ithaca, NY 14853, USA

3. School of Applied and Engineering Physics, Cornell University, Ithaca, NY 14853, USA

4. Laboratory of Atomic and Solid-State Physics, Cornell University, Ithaca, NY 14853, USA

5. Department of Electrical Engineering, University of Notre Dame, Notre Dame, IN 46556, USA

6. Kavli Institute at Cornell for Nanoscale Science, Cornell University, Ithaca, NY 14853, USA

a) ry253@cornell.edu; b) djena@cornell.edu

*03/15/2018*

**Abstract:** We report a systematic study of thickness-dependent superconductivity and carrier transport properties in exfoliated layered 2H-NbS$_2$. Hall-effect measurements reveal 2H-NbS$_2$ in its normal state to be a p-type metal with hole mobility of 1-3 cm$^2$/V·s. The superconducting transition temperature is found to decrease with thickness. We find that the suppression of superconductivity is due to disorder resulting from the incorporation of atmospheric oxygen and a reduced hole density. Cross-section transmission electron microscope (TEM) imaging reveals a chemical change of NbS$_2$ in ambient conditions, resulting in the formation of amorphous oxide layers sandwiching crystalline layered NbS$_2$. Though few-nm-thick 2H-NbS$_2$ completely converts to amorphous oxide in ambient conditions, PMMA encapsulation prevents further chemical change and preserves superconductivity.

The ability to mechanically exfoliate atomically thin films from layered crystals has provided a route to experimentally probe the physics of two-dimensional (2D) semiconductors, metals, and superconductors. The recent discovery of electric field-effect control of superconductivity [1-3], and extremely high in-plane critical magnetic fields in transition metal dichalcogenide (TMD) crystals [2, 4] were both enabled by this ability to isolate ultrathin crystalline layers. When crystals are reduced to few- and single-atomic layers, their electrical and optical properties can change drastically due to changes in the electronic band structure caused by varying interlayer coupling strength and quantum confinement [5-7]. Several studies report that the TMD 2H-NbSe$_2$ exhibits suppressed superconductivity with decreasing thickness [4, 8], while 2H-TaS$_2$ exhibits the opposite trend [9]. Transport in 2H-NbS$_2$ has not yet been well-studied, particularly its superconducting behavior at the ultrathin limit. A bulk superconducting transition temperature $T_c$ ~ 5.4 K has been reported for thick layers of NbS$_2$, which falls between that of bulk TaS$_2$ ($T_c$ ~ 0.5 K) [9] and NbSe$_2$ ($T_c$ ~ 7 K) [4]. Unlike in TaS$_2$ and NbSe$_2$, no evidence for charge density waves in bulk NbS$_2$ has been reported to date [10].

In this work, we have studied the evolution of superconductivity and metallic transport in layered NbS$_2$ flakes with varying film thickness. As with previous studies of NbSe$_2$ [4], we find that ultrathin NbS$_2$ flakes processed in ambient environment show decreasing $T_c$ with decreasing thickness, and they eventually undergo a superconductor-insulator transition when the flake is ~3 nm thin. However, detailed transmission electron microscopy (TEM) and energy-dispersive X-ray spectroscopy (EDS) analysis indicate that the NbS$_2$ flakes prepared in air oxidize during exfoliation, evidenced by the formation of 5-7 nm of niobium oxide sandwiching the crystalline NbS$_2$ layers. Therefore, the transport properties in flakes below 5 nm are significantly affected by NbO$_x$, and the observed transport properties *cannot* be attributed as intrinsic to NbS$_2$. Our findings lead to the simple yet important conclusion that to infer the intrinsic layer-dependent superconductivity and transport properties of single crystal NbS$_2$, one must carefully exclude the naturally-formed oxide layers surrounding the crystalline NbS$_2$.

We fabricated Hall-bar devices (Fig. 1(a) and 1(b)) from ultrathin layers of NbS$_2$ obtained by mechanical exfoliation of bulk crystals (HQ Graphene). The exfoliated flakes were subsequently transferred onto SiO$_2$/Si substrates with a polydimethylsiloxane (PDMS) stamp. Flakes of desirable shape and thickness were first located under an optical microscope by optical contrast and immediately transferred to an atomic force microscope (AFM) for measurement of the thickness. The samples were then covered with a MMA/PMMA electron-beam resist stack. Electron-beam lithography (EBL) and sputtered Cr/Pt (10/50 nm) were used to define metallic contacts to the flakes. After lift-off of the metal layer, the samples were covered with PMMA, followed by an additional EBL step to expose the contact pad regions for subsequent wire-bonding and electrical measurements. Although we have minimized the total exposure time of NbS$_2$ samples to approximately one hour and cover them with PMMA immediately after AFM scanning, the films were found to still unavoidably oxidize.

Cross-section TEM images (Figure 1(d)) of flakes indicate that oxidation occurs on both surfaces of NbS$_2$, and that sufficiently thin flakes may oxidize completely. Figure 1(d) shows a flake with an AFM-measured thickness ($t_{AFM}$) of 9 nm that has developed two amorphous oxide layers of total thickness ~7 nm sandwiching a crystalline region of NbS$_2$. The layer spacing visible in the TEM image is consistent with the 2H polytype crystal structure (Fig. 1(c)). The total thickness measured by TEM ($t_{TEM}$) is approximately 3 nm greater than $t_{AFM}$; this difference arises from further oxidation of NbS$_2$ in the processing steps after the AFM scan (during spinning of electron-beam resist layers and subsequent fabrication steps), up until spinning the PMMA layer that cover the exposed flake. We speculate that the increase in film thickness - from $t_{AFM}$ to $t_{TEM}$ - is due to the incorporation of oxygen in the NbO$_x$ regions. The remaining unoxidized NbS$_2$ layer has a thickness of approximately 5 nm, which is equivalent to 8 atomic layers as determined from the cross-sectional TEM image.

A second, thinner sample (Fig. 1(e)) appears almost completely amorphous, containing only small islands of unoxidized NbS$_2$. For this sample, $t_{AFM} \approx t_{TEM} \approx 5$ nm; the sample had already undergone full oxidation before AFM scans were obtained. TEM analysis of yet another sample with $t_{AFM} \approx 12$ nm (shown in the Supplementary Information) indicates 6-7 nm of NbOx, suggesting that ambient environment exposure was slightly less than the sample in Fig. 1(d) (with 8-9 nm of NbO$_x$).

The processing times for the samples studied here have been kept approximately constant. From this we can estimate based on the AFM and TEM data that the thickness of crystalline NbS$_2$ is 4-5 nm thinner than $t_{AFM}$. Fluctuations in thickness are caused by non-uniform oxidization of the films, as clearly revealed from the TEM images. Furthermore, the EDS analysis shown in the cross-section TEM images of Figs. 1(d) and 1(e) confirms that the amorphous layers in both samples are populated with niobium (Nb), oxygen (O) and sulfur (S). In the $t_{AFM} = 9$ nm sample (Fig. 1(d)), the atomic ratio is Nb:S = 1:2 in the middle of the crystalline region, as expected for stoichiometric NbS$_2$. For the thinner, completely-oxidized sample in Fig. 1(e), the oxygen concentration is higher near the PMMA interface than near the SiO$_2$ interface. In the following, we discuss the transport studies on a series of NbS$_2$ flakes with different thicknesses, thus determining the evolution of superconductivity as a function of the thickness of the unoxidized part of the flakes.

In Figure 2, we demonstrate that the superconducting transition temperature decreases with flake thickness, and our thinnest, fully oxidized sample demonstrates insulating behavior. In Fig. 2(a), we show the measured temperature-dependent resistance $R(T)$ normalized to the room-temperature resistance $R(300\ K)$ for a range of flakes of thicknesses $t_{AFM} = 12, 8, 7, 6, 5, 3$ nm. All these samples show metallic behavior at temperatures exceeding 75 K, characterized by a linear dependence of resistivity on temperature ($R \propto T$) due to phonon-limited scattering [12]. The resistance saturates to a residual value at temperatures below 20 K for all samples, except for the thinnest flake with $t_{AFM} = 3$ nm; impurity scattering is the typical reason for this residual resistivity phenomenon in low-temperature metallic transport. Figure 2(b) shows the normalized resistance from 2 to 10 K, where the resistivity of flakes thicker than 7 nm clearly shows a superconducting transition and the superconducting transition temperature decreases with film thickness. The resistivity of samples of thicknesses 6 nm and 5 nm exhibit trends indicating superconductivity at transition temperatures lower than 2.1 K. In contrast to the superconducting transition of thicker samples, the resistance of the $t_{AFM} = 3$ nm sample undergoes a metal-insulator transition at low temperatures, as clearly shown by the upper curve in Figs. 2(a) and 2(b).

This observation of the transition of transport properties from superconducting to insulating with decreasing film thickness is analogous to early studies reported on layered $NbSe_2$ ultrathin films processed in air [1,11], where the change in behavior was attributed to disorder in the $NbSe_2$ crystal itself. However, more recent reports on atomically thin $NbSe_2$ samples prepared in an inert gas environment confirm that superconductivity still survives to the monolayer $NbSe_2$ limit [3, 4]. As indicated from our TEM and EDS analysis in Fig. 1, we conclude that flakes thinner than 5 nm are likely already fully oxidized and amorphous, therefore, the observed metal-insulator transition behavior observed in the 3-nm sample is due to the presence of non-superconducting amorphous $NbO_x$ and $NbS_x$ that is chemically, electronically, and structurally very distinct from crystalline $NbS_2$ [13, 14].

For a more accurate determination of the superconducting transition temperatures $T_c$, Fig.2(c) shows the Aslamazov-Larkin formula fit as the solid lines to the experimentally measured $R$ vs $T$ [15]. The temperature-dependent resistances are normalized to their corresponding resistances at 8 K. The broadening of the superconducting transition for the thinner flakes is attributed to enhanced thermal fluctuations [2-4, 15, 16] and presence of disorder [1, 10, 15, 16, 17] as captured in the formula. For films thicker than $t_{AFM}=$ 7 nm, the fitted $T_c$ are close to the temperatures at which $R = ½ \times R(8K)$. For the $t_{AFM}=$ 6 nm flake, $T_c$ is slightly smaller than 2.1 K, the lowest temperature that is achieved in our measurement system; the Aslamazov-Larkin fit gives $T_c \sim$ 2 K.

In Fig. 3(a), we summarize the extracted superconducting transition temperatures versus sample thickness $t_{AFM}$ noting that the transition temperatures start deviating from the bulk $T_c$ (~ 5.4 K) at thickness smaller than $t_{AFM} =$ 9 nm. This corresponds to a film with 8-9 monolayers of unoxidized crystalline $NbS_2$ as seen from the TEM image in Fig. 1(a). As the samples get thinner, $T_c$ gradually decreases until it vanishes for $t_{AFM} <$ 5 nm. Open symbols represent the $T_c$ for $t_{AFM} =$ 6 nm sample because for this sample the transition temperature is inferred from the Aslamazov-Larkin fit, unlike the 6 other thicker layers for which the superconducting transition is directly observed.

The observed suppression of superconductivity with decreasing film thicknesses can be a result of enhanced Coulomb scattering caused by charged disorder, which makes it easier to break Cooper pairs for thinner films [1, 3, 17, 18, 19]. To support this hypothesis, the measured residual resistance ratio (RRR), resistivity, and carrier concentration for the series of samples is shown in Figs. 3 (b), 3(c), and 3(d) as a function of thickness. As a conventional method of quantifying the film purity, the RRR defined here as $R(300\ \text{K})/R(8\ \text{K})$ measures the phonon contribution to carrier transport in metallic films [20]; a larger RRR indicates more phonon-dominated near-intrinsic transport properties and hence a less disordered system with less defect and impurity scattering. The RRR shown in Fig. 3(b) clearly follows the trend of suppression of $T_c$, decreasing from ~ 20 in the 15 nm sample to ~ 4 in the 5 nm sample, confirming that thinner samples have more disorder in the electrically active region. In addition, the resistivity of the samples increases with decreasing thickness, as seen in Fig. 3 (c).

Hall-effect measurements reveal that the primary factor for the resistivity increase in thinner samples is not so much due to the degradation in the transport properties, but it is due to the reduction of the mobile carrier concentrations. As shown in Fig. 3(d), the 3D carrier density decreases by almost one order of magnitude, from ~ $2 \times 10^{22}$ cm$^{-3}$ down to ~ $3 \times 10^{21}$ cm$^{-3}$ with the decrease of the sample thickness. Concomitantly, the Hall measurements indicate that NbS$_2$ is a p-type metal with hole mobility in the range of 5-9 cm$^2$/V·s at 6 K and 1-3 cm$^2$/V·s at 300 K in our samples, consistent with previous reports on bulk NbS$_2$ crystals [21]. The reduction of the carrier concentration leads to weakened electron screening and elevated Coulomb interactions. The success of the formation of Cooper pairs (and hence superconductivity) depends on the competition between the attractive force between electrons mediated by virtual exchange of phonons that overcome screened Coulomb repulsion between them. The reduction of mobile carrier density is detrimental to pairing and superconducting transition. This is because with lowered carrier density, the electron screening and Fermi energy are both reduced, which results in the enhanced Coulomb repulsion (hindering the formation of Cooper pairs). Therefore, with these direct measurements of RRR, resistivity, and Hall carrier density, we conclude that the evolution of superconductivity with thickness is dominated by the chemical disorder present in ultrathin $NbS_2$ samples and may not be an intrinsic thickness-dependent property.

In conclusion, by combining TEM, AFM and transport characterization, we have studied the influence of chemical disorder on the low-temperature metallic and superconducting properties of exfoliated $NbS_2$. Upon exposure to the ambient, $NbS_2$ oxidizes rapidly. The oxidation rate is such that between exfoliation and encapsulation the unoxidized crystalline $NbS_2$ is $\sim(5 \pm 1)$ nm thinner than the original thickness measured by AFM. The encapsulation of the samples using a PMMA top layer immediately after fabrication effectively avoids further oxidization of samples. This sample preparation procedure made it possible for us to resolve the layer-dependent superconducting transition temperature in $NbS_2$ flakes. It is found that transition temperatures decrease with decreasing layer number because of the reduction of the mobile hole density in thinner flakes. We emphasize that the observed metal-insulator transition in the thinnest 3-nm sample is *not* the behavior of $NbS_2$; this particular sample is already fully oxidized and has chemically transformed to an amorphous mixture of $NbO_x$ and $NbS_x$. Because of the oxidation, sample preparation in an inert environment is necessary to study the intrinsic behavior of few layer $NbS_2$. A further consequence is that direct epitaxial growth techniques of thin layers such as chemical vapor deposition or molecular beam epitaxy have to be used in a highly controlled chemical environment, and ultrathin crystals of $NbS_2$ have to be capped after epitaxial growth with insulating and chemically stable layers before exposure to the ambient for further characterization of metallic transport and two-dimensional superconductivity in the thinnest samples.

**Acknowledgement.** The measurements performed in this work made use of the Cornell Center for Materials (CCMR) Research Shared Facilities which are supported through the NSF MRSEC program (DMR-1719875). The structure fabrications are realized at the Cornell NanoScale Facility, a member of the National Nanotechnology Coordinated Infrastructure (NNCI), which is supported by the National Science Foundation (ECCS Grant #1542081). This work was partially supported by the CCMR with funding from the NSF MRSEC program (DMR-1719875), and the NSF EFRI 2-DARE grant (Award #1433490) is acknowledged. BTS acknowledges support through an National Science Foundation Graduate Research Fellowship under Grant No. DGE-1650441.

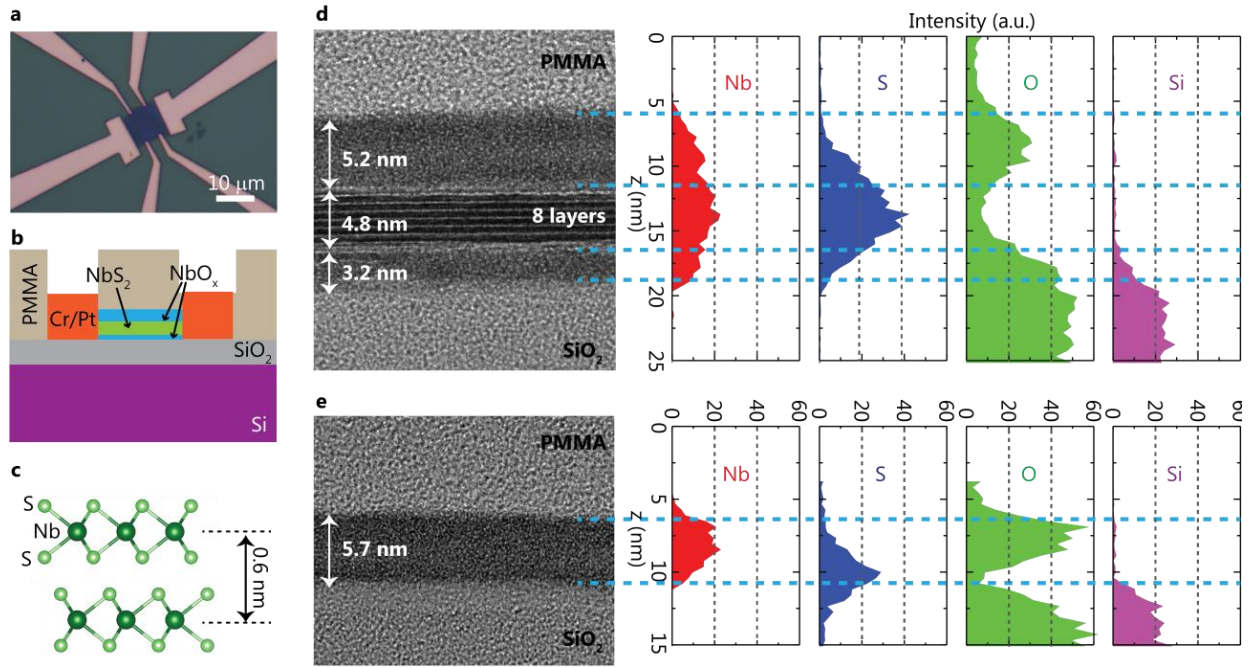

**Figure 1**. (a) Optical microscope image of a fabricated $NbS_2$ sample in a Hall-bar geometry covered with PMMA. The scale bar is 10 μm. (b) Schematic cross-section of fabricated samples. (c) Schematic drawing of the atomic-layer structure of $NbS_2$ with interlayer spacing of 0.6 nm.[22] (d) and (e) Two $NbS_2$ samples of different initial thicknesses, with their TEM cross-section images and EDS analysis along the vertical direction, including the percentages of Nb, S, O, and Si elements. The thicker sample is partially oxidized from the top and the bottom, whereas the thinner sample is completely oxidized. The partially oxidized sample has crystalline layers of $NbS_2$ which are 4-5 nm thinner than the starting flake thickness that is measured by AFM.

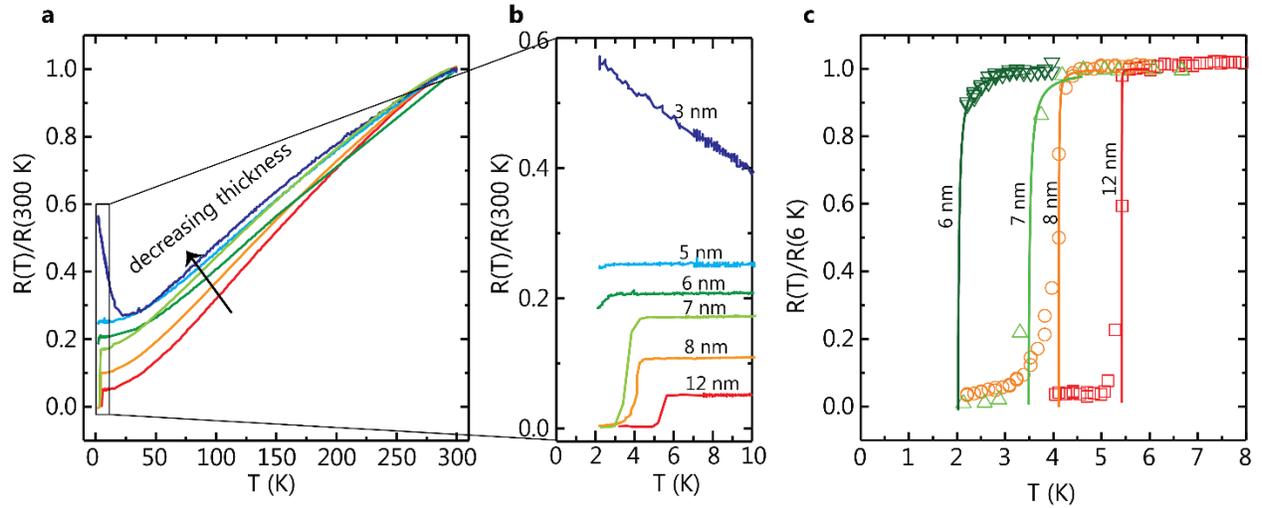

**Figure 2.** (a) Temperature dependence of normalized resistance, $R(T)/R(300\ K)$ for a series of NbS$_2$ samples with different thicknesses showing metallic behavior, a superconducting transition for samples of AFM-determined thicknesses larger than 5 nm, with potentially an onset of the transition for the sample with thickness 3 nm and metal-to-insulator transition for the sample with thickness 3 nm. (b) A zoomed-in view of (a) showing the superconducting transitions clearly. (c) $R(T)/R(8\ K)$ at temperatures close to the superconducting transition. Open symbols are experimental data, and the solid lines are a best-fit to the Aslamazov-Larkin formula [15].

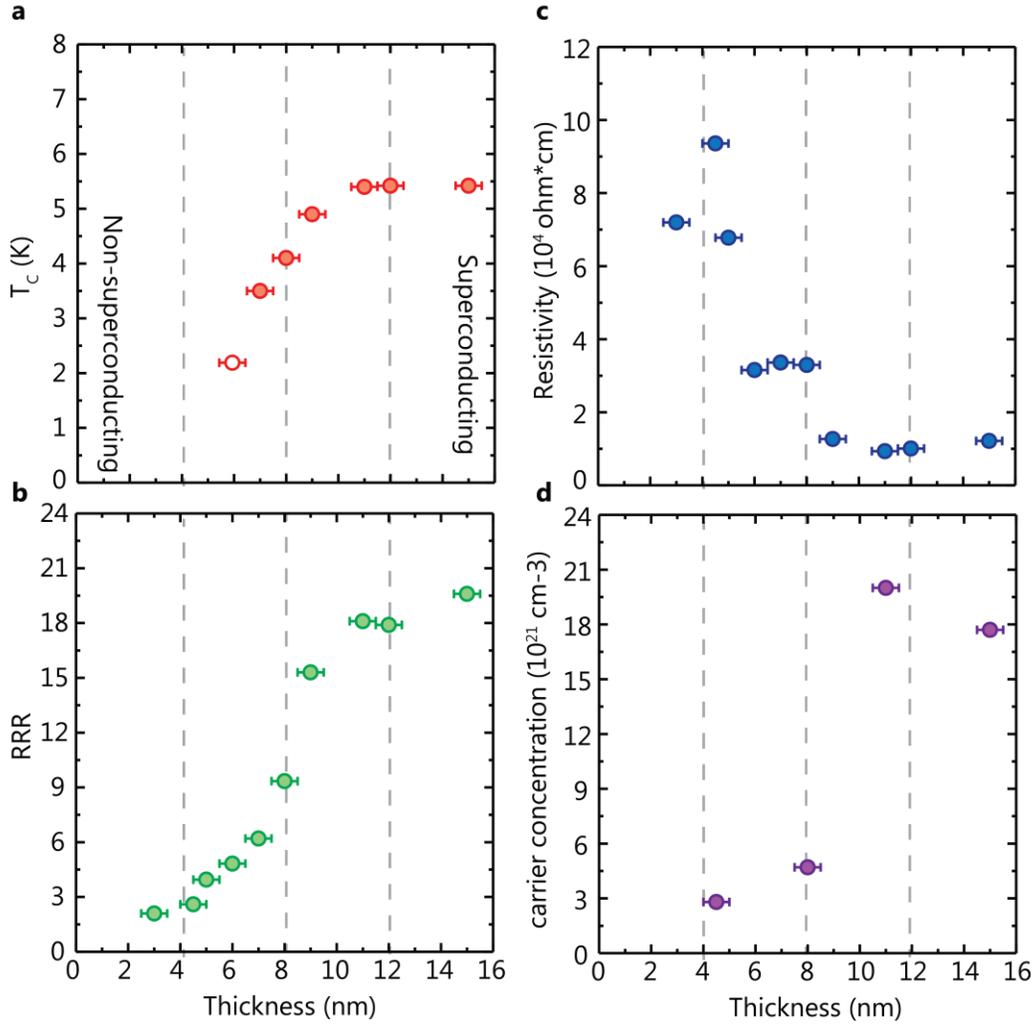

**Figure 3.** (a) The superconducting transition temperature $T_c$, (b) the residual resistivity ratio (RRR), (c) the low-temperature metallic state resistivity, and (d) the Hall-effect carrier concentrations as a function of the film thicknesses of $NbS_2$ as measured by AFM. The dependence of the RRR, resistivity, and the carrier concentrations with thickness are consistent with the reduction of $T_c$ with thickness.

**Supplementary Information for "Evolution of superconductivity in ultrathin NbS$_2$" by Rusen Yan *et al*.**

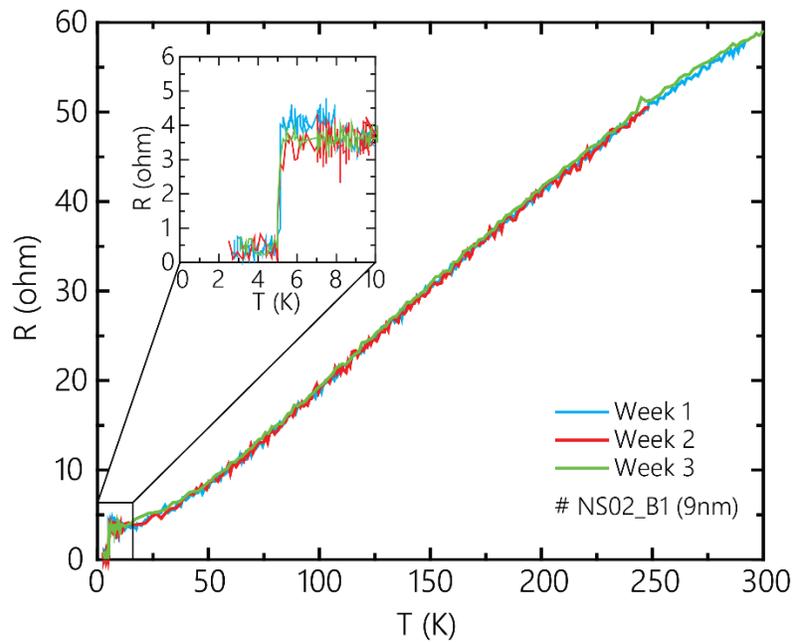

Figure S1. Temperature-dependent resistance measured during the 1st, 2nd and 3rd week after the sample fabrication. The repeatability of these curves indicates the sample quality is maintained for (at least) 3 weeks. All data presented and discussed in this manuscript were collected within two weeks after the samples were prepared.

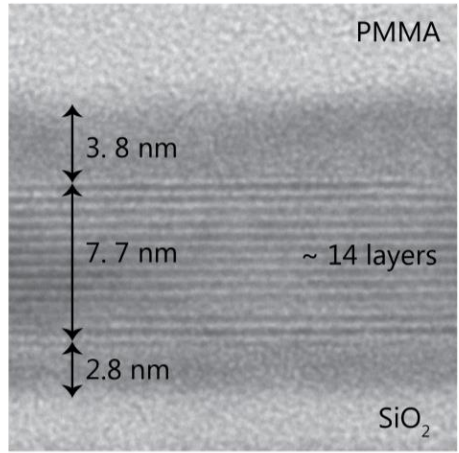

Figure S2. TEM image of an NbS$_2$ sample with $t_{AFM}$ ~ 12 nm. It is seen that the crystalline NbS$_2$ has a thickness of 7.7 nm, which translates to 14 atomic layers. The actual thickness obtained through this TEM image is ~ 13.3 nm.